\documentclass[twoside,12pt]{Latex/Classes/PhDthesisPSnPDF}

\usepackage{tikz}

\usepackage[bookmarks=false]{hyperref}
\usepackage{pgflibraryarrows}
\usepackage{amsmath, amssymb}
\usepackage{graphicx}
\usepackage{color}
\usepackage{listings}
\usepackage{stex-logo}
\usepackage{makeidx}
\usepackage{paralist}
\usepackage[show]{ed}
\lstMakeShortInline[basicstyle=\tt]|
\usepackage{lstomdoc}
\usepackage{url}

\ifpdf  
    \pdfcatalog { /PageMode (/UseOutlines)
                  /OpenAction (fitbh)  }
\fi

\title{Editing Knowledge in Large Mathematical Corpora. \\ A case
  study with \\ Semantic \LaTeX{} (\stex{}).}

\ifpdf
  \author{\href{mailto:c.jucovschi@jacobs-university.de}{Constantin Jucovschi}}
  \collegeordept{\href{http://www.eecs.iu-bremen.de/wiki/index.php/Smart_Systems}{Computer Science}}
  \university{\href{http://www.jacobs-university.de}{Jacobs University Bremen}}

  \crest{\includegraphics[width=4cm]{logo.png}}
  \supervisor{Prof.\ Dr.\ Michael Kohlhase}

\else
  \author{Constantin Jucovschi}
  \collegeordept{Computer Science}
  \university{Jacobs University Bremen}
  \crest{\includegraphics[width=4cm]{logo.png}}
\fi

\degree{Master of Science}
\degreedate{2010 August}

\newcommand{\stexide}{sTeXIDE}
\newcommand{\omdoc}{OMDoc}
\newcommand{\casl}{CASL}
\newcommand{\xtext}{xText}
\newcommand{\unicode}{unicode}
\newcommand{\tbs}{\textbackslash}
\newcommand{\gencs}{GenCS}
\newcommand{\latexml}{\LaTeX ML}
\newcommand{\latexmlpost}{\LaTeX MLPost}

\newcommand*{\person}[1]{\textsc{#1}}

\makeindex

\begin{document}


\renewcommand\baselinestretch{1.2}


\maketitle  










\begin{declaration}        

The research subsumed in this thesis has been conducted under the
supervision of Prof. Dr. Michael Kohlhase from Jacobs University
Bremen. All material presented in this Master Thesis is my own, unless
specifically stated. 

I, Constantin Jucovschi, hereby declare, that, to the best of my knowledge, the
research presented in this Master Thesis contains original and
independent results, and it has not been submitted elsewhere for the
conferral of a degree. 

\vspace{10mm}

Constantin Jucovschi, \\
Bremen, August 23rd, 2010

\end{declaration}


\begin{acknowledgements}      

I would like to thank my supervisor, Prof. Dr. Michael
Kohlhase, for the opportunity to do great research with him, for the
guidance I have received for the past 3 years and for his genuine
interest in finding the topic I ``tick for''. I want to thank him for
the great deal of support and understanding he showed when it came to
more personal matters. I could not be happier to have Prof. Dr. Kohlhase
as supervisor for my doctoral studies. 

I would also like to thank the members of the KWARC group for creating
such a motivating and in the same time pleasant research
environment. Thank you for all the questions, hints and discussions --
they helped a lot; and special thanks for the ability to interact with you
also outside academic environment. 

I would like to express my appreciation and gratitude to my
family, for their long-term engagement to support and encourage me in
all my beginnings. I owe most of my success to you.
\end{acknowledgements}




\begin{abstracts}        
Before we can get the whole potential of employing computers in the
process of managing mathematical `knowledge', we have to
convert informal knowledge into machine-oriented  representations. How
exactly to support this process so that it becomes as
effortless as possible is one of the main unsolved problems of
Mathematical Knowledge 
Management. 

Two independent projects in formalization of mathematical content
showed that many of the time consuming tasks could be significantly
reduced if adequate tool support were available. It was also
established that similar tasks are typical for object oriented
languages and that they are to a large extent solved by Integrated
Development Environments (IDE). 

This thesis starts by analyzing the opportunities where formalization
process can benefit from software support. A list of research
questions is compiled along with a set of software requirements which
are then used for developing a new IDE for the semantic \TeX{}
(\stex{}) format. The result of the current research is that, indeed,
IDEs can be very useful in the process of formalization and presents a set of
best practices for implementing such IDEs.
\end{abstracts}




\setcounter{secnumdepth}{3} 
\setcounter{tocdepth}{3}    
\tableofcontents            




\markboth{\MakeUppercase{\nomname}}{\MakeUppercase{\nomname}}


\nomenclature{AST}{Abstract Syntax Tree} 


\begin{multicols}{2} 
\begin{footnotesize} 

\printnomenclature[1.5cm] 
\label{nom} 

\end{footnotesize}
\end{multicols}


\mainmatter

\renewcommand{\chaptername}{} 




\chapter{Introduction}
\begin{quotation}
{ \it
 I used to come up to my study, and start trying to find patterns. I
 tried doing calculations which explain some little piece of
 mathematics. I tried to fit it in with some previous broad conceptual
 understanding of some part of mathematics that would clarify the
 particular problem I was thinking about. Sometimes that would involve
 going and looking it up in a book to see how it's done
 there. Sometimes it was a question of modifying things a bit, doing a
 little extra calculation. And sometimes I realized that nothing that
 had ever been done before was any use at all. Then I just had to find
 something completely new; it's a mystery where that comes from. }
\begin{flushright} Solving Fermat, \\Sir Andrew John Wiles \cite{wiles_fermat:web} \end{flushright}
\end{quotation}

One of the distinctive features of Mathematics is its intrinsic 
dependence on existing knowledge \cite{sojka2010dml, bouche2010digital}. 
As part of everyday scientific life, mathematicians use books,
journals, the internet etc., to look up formulas, methods and tools and
use them to discover new patterns, formulate conjectures and 
establish truths. Reporting results back to the community by writing
papers, participating in conferences or even blogging are ways to
get community feedback as well as recognition. Hence consulting and
contributing to sources of mathematical knowledge is a vital part of a
mathematician's research life. 

The most recent estimate for the volume of mathematical knowledge
produced each year is about 3 million pages
\cite{bouche2010digital}. Obviously there is no chance for a person to
even read (not to mention digest), such volumes of information. Of course 
not all 3 million pages are relevant for a particular researcher, thus
mathematicians select only certain conferences or journals 
which they follow closely. A serious drawback of such an approach is that
mathematical results discovered in some branch of mathematics  
stay unknown in other communities solving similar problems. There is
no simple way of solving this problem because in many cases, even 
a mathematician familiar with both topics might not see how problems are
related, as the conceptual mapping involved may be not-trivial. Yet,
I conjecture that there is a lot of potential to be uncovered by 
getting better computer support in structuring mathematical knowledge and
searching it for relevant documents. 

Given the importance of mathematical knowledge for the whole
scientific community, it is not a coincidence that there is an
emerging research field known as Mathematical Knowledge Management
(MKM) which defines its objective ``{\it to develop new and
  better ways of managing mathematical knowledge using sophisticated
  software tools}''. As the topic of this thesis fits well in this
objective and often refers to the results achieved in the MKM
community, I will dedicate the section \ref{sec:intro:mkm} to
introduce the main research directions and identify which of them are
important to current research. In section \ref{sec:intro:udml}, one of
the main long term objectives of MKM is presented, namely the creation
of a Universal Digital Mathematical Library (UDML). In sections 
\ref{intro:sec:structure} and \ref{sec:into:degrees_formality} I will
present what paradigms are used today to work with mathematics and
identify some weak points which could hinder the successful
implementation of UDML. I will finish this chapter (section
\ref{sec:intro:flexiforms}) with presenting changes to current ways of
thinking about mathematical structures and formality which alleviate
the weak points identified in section \ref{sec:into:degrees_formality}. 

\section{Mathematical Knowledge Management}
\label{sec:intro:mkm} \index{Mathematical Knowledge Management (MKM)}
In this section I would like to specify the scope of the MKM research
field by giving an overview of the challenges it is trying to 
address. As the set of challenges is quite big, only the questions
relevant to current research will be mentioned. In the next section
(\ref{sec:intro:udml}) I will also introduce the ``Grand
Challenge'' of MKM which is a project integrating all the aspects of MKM. 
 
There are 4 levels at which the subject of Mathematical Knowledge
Management can be addressed:
\begin{compactenum}
  \item [\textbf{document level}] --- addresses low level document
    issues like format, level of formality, context and
    representation. Questions relevant for current research are:
    \begin{compactenum}
      \item [\textbf{D1}] What software support is needed to convert informal
        mathematical documents to formal?
      \item [\textbf{D2}] When do benefits of formalizing mathematical knowledge
        outweigh the costs?
      \item [\textbf{D3}] How should be context of mathematical knowledge expressed?
    \end{compactenum}
  \item [\textbf{organization level}] --- concentrates on 
    knowledge reuse, inter-document linking,  dealing with
    theoretically infinite size of  mathematical knowledge. On this
    level I am interested in the questions:
    \begin{compactenum}
      \item [\textbf{O1}] how should formal as well as informal documents be linked
        to avoid redundancy?
      \item [\textbf{O2}] what tools are needed to deal efficiently with highly
        interconnected structures?
    \end{compactenum}
  \item [\textbf{dissemination level}] --- deals with administrative
    questions like certification of knowledge, effective dissemination
    of mathematical knowledge, ownership of data. These questions are
    irrelevant for the scope of current thesis. 
  \item [\textbf{end-user tools level}] --- establishes end-user
    requirement for tools and services to efficiently work with MKM
    corpora. Relevant questions for this category are:
    \begin{compactenum}
      \item [\textbf{E1}] possibility of math oriented semantic searches?
    \end{compactenum}
\end{compactenum}

\section{Universal Digital Mathematics Library}
\label{sec:intro:udml} \index{Universal Digital Mathematics Library (UDML)}
Building a Universal Digital Mathematics Library (UDML) is the ``Grand
Challenge'' of the Mathematical Knowledge Management (MKM) community
according to \person{Farmer} \cite{Farmer:mkm05}. Note that the UDML
is a library rather than an 
archive of mathematical knowledge. The difference is that UDML is 
expected to provide a much larger range of interaction 
and would be continuously reorganized as new discoveries and
connections are made. Here is a summary of requirements as seen by
\person{Farmer} for constructing UDML 
\begin{compactenum}
\item [\textbf{creation}] --- UDML should be constructed in an open 
  collaborative way and be accessible through the internet.
\item [\textbf{structure}] --- mathematical knowledge would be very
  structured and would contain highly interconnected mixture of
  axiomatic, algorithmic, diagrammatic and other types of mathematical
  knowledge.
\item [\textbf{maintenance}] --- our understanding of mathematics
  changes as new concepts and generalizations are introduced. It is
  important to be able to change and adapt knowledge structure to
  accommodate the state of the art in mathematics.
\item [\textbf{correctness}] --- mathematical content would carry a
  certification of correctness.
\item [\textbf{tools}] --- UDML would provide a set of tools for
  exploring, editing and searching mathematical content.
\end{compactenum} 






\section{Structure of Mathematics}
\label{intro:sec:structure}
In many ways, mathematics can be seen as a complex network of
interconnected and mutually supporting knowledge items. This statement
is confirmed by several major independent efforts to structure
mathematical knowledge into reusable components. I will emphasize the
difference between approaches by comparing how the definition of a
simple mathematical object, such as the concept of a monoid, is
presented. First, let us consider a typical definition of monoids.

\begin{figure}[ht]
\begin{quotation}
\it A monoid is a tupple (M, *) where $M$ is a set and $*$ is a
binary operator; such that
\begin{enumerate}
\item closure: $\forall a, b \in M \Rightarrow a * b \in M$
\item associativity: (a * b) * c = a * (b * c)
\item identity element: $\exists e \in M, a * e = e * a = a$
\end{enumerate}
\end{quotation}
\caption{Typical definition of monoids}
\label{fig:intro:monoid-typical}
\end{figure}

One of the most influential attempts of restructuring mathematics in reusable
components is the series of books ``Elements of Mathematics'',
written by a group of french mathematicians working under the
pseudonym \person{Nicolas Bourbaki} \cite{Bourbaki:a68}. The
motivation of the group was to break down existing mathematical
knowledge to its core elements. So just like in chemistry, where
molecules can be identified by structures connecting basic chemical
elements, \person{Bourbaki} tried to describe known mathematical
objects in terms of a minimal set of elementary mathematical elements
i.e. axioms. The project run for more then 50 years, time in which 9
highly rigorous books covering core ares of modern mathematics were
written. Consider the definition of monoids as given by
\person{Bourbaki} \cite{Bourbaki:a74}
\begin{quotation}
\it Definition 2. A magma with an identity element is called a unital
magma [...] An associative unital magma is called a monoid. 
\end{quotation}
Compared to the definition in figure \ref{fig:intro:monoid-typical}, the
definition above is very compact because it reuses the structure of a
magma and only tells how to extend it get a monoid object. 

Even though it is widely agreed that the efforts of the
\person{Bourbaki} group are still valuable, they are only intended for
the human user. The reason for it is the implicit structure of objects
which becomes clear only after careful reading of the text. 
With the advance of information technologies, new demand for
working and structuring mathematics appeared (e.g. from automatic theorem
provers) and gave rise to new efforts like Mizar \cite{mizar}, DLMF
\cite{Loz:DLMF} etc. These, however, also concentrated on making the
structure computer understandable and introduced notions like little
theories \cite{FaGu:lt92}, development graphs\index{Development graphs} \cite{MAH06} and
culminating in modular and web-scalable \cite{KohRabZho:tmlmrsca10}
representations of mathematical knowledge. 

\newcommand{\cn}[1]{#1}
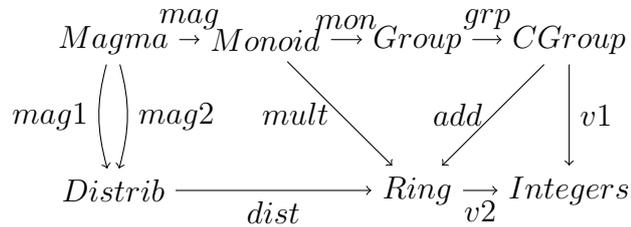
\begin{figure}
\centering
\begin{tikzpicture}
\node (M)  at (0,0) {$\cn{Magma}$};
\node (Mo) at (2,0) {$\cn{Monoid}$};
\node (G)  at (4,0) {$\cn{Group}$};
\node (GA) at (6,0) {$\cn{CGroup}$};
\node (R)  at (4,-2){$\cn{Ring}$};
\node (D)  at (0,-2){$\cn{Distrib}$};
\node (I)  at (6,-2){$\cn{Integers}$};
\draw[->](M)  --node[above] {$\cn{mag}$} (Mo);
\draw[->](Mo) --node[above] {$\cn{mon}$} (G);
\draw[->](G)  --node[above] {$\cn{grp}$} (GA);
\draw[->](GA) --node[left] {$\cn{add}$} (R);
\draw[->](Mo) --node[left] {$\cn{mult}$} (R);
\draw[->](M) to[out=-105,in=105] node[left] {$\cn{mag1}$} (D);
\draw[->](M) to[out=-75,in=75] node[right] {$\cn{mag2}$}(D);
\draw[->](D)  --node[below] {$\cn{dist}$} (R);
\draw[->](GA)  --node[right] {$\cn{v1}$} (I);
\draw[->](R)  --node[below] {$\cn{v2}$} (I);
\end{tikzpicture}\vspace*{-1em}
\caption{Algebraic Hierarchy}\label{fig:mmttnt:algebra}\vspace*{-2em}
\end{figure}

The current state of the art in representing formal mathematical
knowledge is based on the concept of mathematical theories and theory
graphs. To the best knowledge of the author, the MMT approach
\cite{Rabe:MMTLanguageSystem09} of representing mathematical theories
and relationship between them presents the latest development in
this field. Hence when speaking about mathematical theories, theory
morphisms and theory graphs, I mean the concepts with the same names
from the MMT approach. 

As the name already suggests, theory graphs are directed graphs where nodes
represent theories, edges --- theory morphisms (see figure
\ref{fig:mmttnt:algebra}). A mathematical theory consists of a 
\begin{compactenum}
\item[\textbf{collection of symbol declarations}] --- names for
  mathematical objects that are particular for that theory. In the
  case of monoids, we have 3 symbols ($M, e, *$) corresponding to the ones
  in definition \ref{fig:intro:monoid-typical}. 
\item[\textbf{axioms}] --- logical formulas which state laws governing
  the objects described by the theory. Monoids theory only needs the
  {\it identity element } and {\it assiciativity } axioms. 
\end{compactenum}

A theory morphism from theory $A$ to $B$, identifies symbols from
theory $A$  with symbols in the destination theory $B$. Using this
mapping, one can ``transport'' axioms as well as theorems from one
theory to another (see \cite{Rabe:MMTLanguageSystem09} for detailed
description). For example, the morphism between the Magma theory and
Monoid theory in figure \ref{fig:mmttnt:algebra} makes sure that the
closure axiom is part of the Monoid theory without being explicitly
added to the list of axioms. The advantage of representing mathematical knowledge by using
theory graphs is the explicit structure which potentially brings a lot
of computer support for creating, managing and visualizing
mathematical knowledge. 

Comparing the representation of the monoid object as given by
\person{Bourbaki} with that of theory graphs, we see that they have a
lot in common. Namely, they both reuse the structure of the magma
object and add the {\it associativity} and {\it identity element}
axioms. The reuse mechanism employed by both approaches very much
resembles the inheritance paradigme from object oriented
programming. This similarity motivated our approach of using IDEs for
the authoring process of mathematical documents.

\section{Dimensions of Formality} \index{MMT} \index{Dimensions of Formality}
\label{sec:into:degrees_formality}
In mathematics, a document is considered formal if it supports
syntax-driven reasoning processes e.g. by using it into an automatic
theorem  prover. In the section 
\ref{intro:sec:structure}, I presented three examples on how the
concept of a monoid can be defined. All three examples are formal
because one can easily derive an equivalent set of axioms and derivation
rules which lead to syntax-driven reasoning. However, when comparing the
definitions from \person{Bourbaki} and those described by MMT theory
graphs, one has the feeling that MMT theory graphs are somewhat more
formal. Indeed they are, however not in the mathematical sense of
formality defined above. Theory graphs are more formal from the
content organization point of view i.e. one can do syntax-driven
reasoning about the structure of the content e.g. reason about reuse, symbol
visibility etc. -- things one cannot do with \person{Bourbaki}
theories. Hence MMT theory graphs are formal in both mathematical
sense as well as content structure sense. This realization gives us an
idea that there are different dimensions of formality.

One might be tempted to think that the dimensions of formality are
orthogonal. While this might hold for particular pairs of
formalization, it is usual that formalizing a document in one
dimension also formalizes the other dimensions. For example, the
\person{Bourbaki} formalization of mathematics induces an implicit
content structure -- that is why the \person{Bourbaki} definition of a
monoid is so similar to the one in MMT (e.g. both extend the magma
object). In fact, if Natural Language Processing (NLP) tools were
powerful enough, one could automatically translate \person{Bourbaki}
formalizations to MMT. Hence the formalizations are ``NLP'' distance
apart, which is not much. 

One big advantage of working with formalized documents is the
possibility of getting computer support. Obviously, the content should
be available in some computer understandable format but this is
no longer an issue if the formalization process is supported by
software tools from the very beginning. That is why nowadays,
developing a MMT based formal corpora is not harder then writing 
\person{Bourbaki} style formalizations. On the contrary,
computer tools proved to be extremely useful in the process of
formalization as they could be used to perform type checking to spot
inconsistencies and automatically find formal proofs. 

Still, the process of mathematical content formalization is difficult, even
when supported by modern software products. One of the main reasons
for it is the fragility of the ``formal'' 
state of a document. Namely, by introducing an informal but natural
for mathematicians term like ``obviously'', suddenly brings the
document to an informal state. Even phrases like ``similar to case
X'', which specify the approximate structure of the formal object they
replace, is usually impossible for the computer to formalize. Humans
have the ability of gracefully handling both of the aforementioned
phrases. This is due to the much higher level of understanding of the
mathematical objects they work with.

\section{Flexiformalization} \index{Flexiform} \index{Formalization}
\label{sec:intro:flexiforms}

As already mentioned in the previous section \ref{sec:into:degrees_formality}, 
formalization of mathematical objects is a very demanding task. It
seems plausible that by formalizing mathematical content, one
simultaneously does an implicit content structure formalization. No
wonder that the process of formalization is so unintuitive and
cumbersome. This is similar to an attempt of jumping over several
stairs (of formality) in one leap. As only a handful of people are
even able to do this kind of a jump in formality, I cannot expect
large communities of contributors to appear and work towards
formalization of mathematics.  

Another point I mentioned in section \ref{sec:into:degrees_formality}
is the formal state of a document is very fragile and can be easily 
broken. This is, however, an artificially built limitation by the
classical definition of a formal object. Namely, in the classical
sense, one expects that a formalized object can be directly used in a 
theorem prover -- the classical consumer of formalized
mathematics. If the aim of formalization is to construct proofs,
process which requires 100\% formalization of all the objects
properties, then one really needs to work with fully formalized
objects. But as I already mentioned in sections \ref{sec:intro:mkm}
and \ref{sec:intro:udml}, the current most stringent MKM challenges don't
even mention the problem of automatic proof finding. It is not even
clear that by having a fully formalized corpus of mathematical
knowledge would solve all of the MKM problems.

An idea proposed by \person{Michael Kohlhase}
\cite{KohKohLan:tcmff10}, is to coin a new, more
flexible term for the existing notion of formal, namely:
\begin{quotation}
\it  We will use the word \textbf{flexiform} as an adjective to
describe the fact that a representation is of \textbf{flexible
  formality}, i.e., can contain \textbf{informal} (i.e., appealing to 
a human reader) and \textbf{formal} (i.e., supporting syntax-driven
reasoning processes) components or both.
\end{quotation}
From the computer perspective, this definition reads: {\it you (the
computer) will no longer have the luxury of understanding
everything. Instead, you'll have to deal with many things you don't
understand but now and then will find some things you do
understand}. Doesn't that sound similar to the way humans deal with
reality? 

One of the biggest advantages of flexiformal documents is that instead
of the formal/informal dichotomy, one gets a whole spectra of
formality. With flexiforms, one can develop formalizations in a
progressive fashion e.g. (from  \cite{KohKohLan:tcmff10}) 
\begin{compactenum}
  \item [{\it i}] an informal proof sketch on a blackboard, and
  \item [{\it ii}] a high-level run-through of the essentials of a
    proof in a colloquium talk, and
  \item [{\it iii}] the author's notes that contain all the details that are glossed over in
  \item [{\it iv}] a fully rigorous proof published in a journal, which may lead to
  \item [{\it v}] a mechanical verification of the proof in a proof checker.
\end{compactenum}
It is quite clear that formalization, which in our example means
transforming $i \to v$, is much harder to achieve then going from one
level of flexiformalization to the next. The gained simplicity
increases a lot the number of potential contributors to
flexiformalization efforts. 



\chapter{Aims of the project} 

\section{Scope}
Due to the rigid definition of formality (see
\ref{sec:into:degrees_formality}), the biggest majority of
existing mathematical documents are classified as informal. And yet, the
amount research efforts and developed tools for formalized
mathematics is much bigger then for informal documents. Hence, as soon
as a document is classified as informal, one suddenly gets very limited
tool support. In the current thesis, I am interested in working
with informal documents and supporting the process of
flexiformalization. By that, I hope to reduce the discontinuity of
tool support between formal and informal documents. 

There are several formats supporting flexible formalization of
documents e.g. MathDox \cite{ccb:MDMDonW}, MathLang
\cite{KWZ:CmtiM-long}, \omdoc{} \cite{Kohlhase:OMDoc1.6spec}. In the
current research, I will concentrate on the \LaTeX{}-based front-end for \omdoc{}
format called \stex{} to achieve flexiformalization. This choice was
mostly influenced by the abundance of informal \LaTeX{} documents as
well as the practical applications which motivated current research (see
\ref{sec:aims:motiv}). Still, I believe that most of the results and
findings are independent of the flexiformalization format. 


The current chapter will progress as follows. In section
\ref{sec:aims:motiv}, I will present the challenges of previous efforts
in flexiformalization of mathematical documents. Then, in section
\ref{sec:aims:aims}, I will compile an extensive list aims for the
current research by combining the general MKM questions presented in
section \ref{sec:intro:mkm} and previous experience presented in
\ref{sec:aims:motiv}.

\section{Motivation}
\label{sec:aims:motiv}
Before we can get the whole potential of employing computers in the
process of managing mathematical `knowledge' --- i.e. reuse and
restructure it, adapt its presentation to new situations,
semi-automatically prove conjectures, search it for theorems
applicable to a given problem, or conjecture representation theorems,
we have to convert informal knowledge into machine-oriented 
representations. How exactly to support this formalization process so
that it becomes as effortless as possible is one of the main unsolved
problems of MKM. \index{LaTeX} \index{sTeX}

Currently most mathematical knowledge is available in the form of
{\LaTeX}-encoded documents. To tap this reservoir \person{Kohlhase}
developed the  \stex~\cite{Kohlhase:ulsmf08,sTex:web} format, a
variant of {\LaTeX} that is geared towards marking up the semantic
structure underlying a mathematical document. 

In the last years, \stex{} has been used in two larger case
studies. In the first one, \person{Kohlhase} has accumulated a
large corpus of teaching materials, comprising more than 2,000 slides,
about 800 homework problems, and hundreds of pages of course notes,
all written in \stex{}. The material covers a general first-year
introduction to computer science, graduate lectures on logics, and
research talks on mathematical knowledge management. The second case
study consists of a corpus of semi-formal documents developed in the
course of a verification and SIL3-certification of a software module
for safety zone computations \index{LaTeXML}
\cite{KohKohLan:difcsmse10,KohKohLan:ssffld10}. In both cases, it was
very useful and important that \stex documents can be transformed into the 
XML-based \omdoc~\cite{Kohlhase:OMDoc1.2} by the {\latexml}
system~\cite{Miller:latexml}, see~\cite{KohKohLan:difcsmse10}
and~\cite{DKLRZ:PubMathLectNotLinkedData10} for a discussion on the
MKM services afforded by this. 

\index{OMDoc}
These case studies have confirmed that writing \stex is {\emph{much}}
less tedious than writing \omdoc{} directly. Particularly useful was the
possibility of using the \stex-generated PDF for proofreading the text
part of documents. Nevertheless serious usability problems
remain. They come from three sources: 
\begin{compactenum}
\item[\textbf{P}1] installation of the (relatively heavyweight)
  transformation system (with dependencies on \texttt{perl},
  \texttt{libXML2}, {\LaTeX}, the \stex packages),
\item[\textbf{P}2] the fact that \stex supports an object-oriented
  style of writing mathematics, and
\item[\textbf{P}3] the size of the collections which make it difficult
  to find reusable components.
\end{compactenum}
The documents in the first (educational) corpus were mainly authored
directly in \stex{} via a text editor (\texttt{emacs} +
\texttt{AUCTeX} mode). This was serviceable for the author, who had a good
recollection of the about 2200 names of semantic macros he 
had declared, but presented a very steep learning curve for other
authors (e.g. teaching assistants) to join. The software engineering
case study was a post-mortem formalization of existing (informal)
{\LaTeX} documents.  Here, installation problems and refactoring
existing {\LaTeX} markup into more semantic \stex{} markup presented the
main problems. 
\index{Integrated Development Environment}
\index{Eclipse}
For programs, similar authoring and source management problems are tackled by
Integrated Development Environments (IDEs) like
\textsc{Eclipse}~\cite{Eclipse:web}, which integrate support for
finding reusable functions, refactoring, documentation, build
management, and version control into a convenient editing
environment. In many ways, \stex shares more properties with
programming languages like \textsc{Java} than with conventional
document formats, in particular, with respect to the three problem
sources mentioned above 
\begin{compactenum}
\item[\textbf{S}1] both require a build step (compiling \textsc{Java}
  and formatting/transforming \stex{} into PDF/\omdoc{}),
\item[\textbf{S}2] both favor an object-oriented organization of
  materials, which allows to 
\item[\textbf{S}3] build up large collections of re-usable components
\end{compactenum}
\index{sTeXIDE}
To take advantage of the solutions found for these problems by
software engineering, we have decided to developed the \stexide{}
integrated authoring environment for \stex-based representations of
mathematical knowledge.

\section{Aims}
\label{sec:aims:aims}
In section \ref{sec:intro:mkm} I already presented a list of
MKM questions relevant to current research. The questions are,
however, too general to be answered in this thesis. Hence I
will restrict the scope of the questions by adapting them to the 
idea of using an Integrated Development Environment to solve the
challenges described in section \ref{sec:aims:motiv}. In the following
list of revised questions, I will use the notation e.g. D1 $\to$ A1 to
show that question D1 from section \ref{sec:intro:mkm} is transformed to
research question A1. 
\begin{compactenum}
\item[\textbf{D1 $\to$ A1}] What features can an IDE provide to make the
  process of converting informal mathematical documents to formal as
  easy as possible? 
\item[\textbf{D2 $\to$ A2}] Due to time restrictions, it was not
  feasible to perform a usability study for getting a rough estimate
  of the benefits and IDE can provide. A fist step towards answering
  the question would be estimating costs. Hence our research
  question A2 is ``What are the costs of developing an IDE for
  languages similar to \stex{} and extending it with new features?''
\item[\textbf{D3 $\to$ A3}] How can an IDE help identifying
  contextual information? 
\item[\textbf{O1 $\to$ A4}] What features can assist the user in
  creating reusable content to avoiding redundancy?
\item[\textbf{O2 $\to$ A5}] What tools can an IDE provide to ease
  navigation through highly interconnected structures?
\item[\textbf{E1 $\to$ A6}] What are the perspectives of math oriented
  semantic search in an IDE?
\end{compactenum}


\chapter{State of the Art} 
In this chapter I will introduce the main technologies relevant
to the current research and briefly summarize them. In the first section,
I will present the \omdoc{} markup language as well as the semantic
\LaTeX{} (\stex{}) language used as front-end language to generate
\omdoc{} documents. In the second section, I will introduce the tools
and frameworks we used to accomplish the aims mentioned in section
\ref{sec:aims:aims}. 

\section{Mathematical Knowledge Management}
\subsection{OMDoc}
\index{OMDoc}
As defined by \person{Kohlhase} \cite{Kohlhase:OMDoc1.2}, ``{\it The OMDoc
  (Open Mathematical Documents) format is a content markup scheme for
  (collections of) mathematical documents...}''. The key element of
this definition is that \omdoc{} is a content markup scheme. In
comparison to presentational markup schemes (like HTML) which change
the way a document is rendered, \omdoc{} concentrates solely on making
meaning of mathematical structures and relationships between them
explicit. A useful feature of the format is that it also integrates
well with informal knowledge and hence represents a natural candidate for
working with flexiform documents. \index{Flexiform}

The \omdoc{} format acts on 3 levels of the document structure. On the:
\begin{compactenum}
  \item [\textbf{object level}] -- it represents
    content of mathematical formulae in one of the established
    standards OpenMath \cite{AbbLeeStr:ocmicakn98} or Content-MathML
    \cite{CarlisleEd:MathML3}. These provide content markup to
    represent formulae structure as well as context markup to link
    formula symbols to already defined entities.
  \item [\textbf{statement level}] provides original markup for making
    structure of mathematical statements (e.g. axioms, definitions,
    examples etc.) explicit. It also provides ways to specify
    contextual information, for example, one can indicate the
    definition that an example is illustrating. Then
  \item [\textbf{theory level}] supplies original markup for
    clustering set of statements into theories, and specifies
    relations between theories by morphisms.  
\end{compactenum}
\index{MMT}
It is easy to notice the similar principles of the \omdoc{} and
MMT formats e.g. in the organization of content into theories and theory
morphisms we recognize the reuse patterns of MMT. The difference is
that \omdoc{} can handle any flexiform document and to some extent
formalized MMT documents. In fact, the next version of \omdoc{} will
be based on MMT and hence be able to fully cover the formal part of
mathematics as well.

\subsection{\protect\stex{}}
\index{sTeX} \index{LaTeX}
The main concept in \stex is that of a ``{\emph{semantic macro}}'',
i.e. a {\TeX} command sequence $\mathcal{S}$ that represents a
meaningful (mathematical) concept or object $\mathcal{O}$: the {\TeX}
formatter will expand $\mathcal{S}$ to the presentation of
$\mathcal{O}$. For instance, the command sequence |\positiveReals| is
a semantic macro that represents a mathematical symbol --- the set
$\mathbb{R}^+$ of positive real numbers. While the use of semantic
macros is generally considered a good markup practice for scientific
documents\footnote{For example, because they allow adapting notation
  by macro redefinition and thus increase reusability.}, regular
{\TeX/\LaTeX} does not offer any infrastructural support for
this. \stex does just this by adopting a semantic, ``object-oriented''
approach to semantic macros by grouping them into ``modules'', which
are linked by an ``imports'' relation.  To get a better intuition,
consider the example in listing~\ref{lst:stex-ex}.

\begin{lstlisting}[label=lst:stex-ex,caption=An \protect\stex module for Real Numbers,escapechar=|,language=sTeX,numbers=none]
\begin{module}[id=reals]
  \importmodule[../background/sets]{sets}
  \symdef{Reals}{\mathcal{R}}
  \symdef{greater}[2]{#1>#2}
  \symdef{positiveReals}{\Reals^+}
  \begin{definition}[id=posreals.def,title=Positive Real Numbers]
    The set $\positiveReals$ is the set of $\inset{x}\Reals$ such that $\greater{x}0$
  \end{definition}
  |\ldots|
\end{module}
\end{lstlisting}
which would be formatted to

\begin{quote}\hrule\vspace*{.3em}
  \textbf{Definition} 2.1 (Positive Real Numbers): \\
  The set $\mathbb{R}^+$ is the set of $x\in\mathbb{R}$ such that
  $x>0$\vspace*{.3em}\hrule 
\end{quote}

Note that the markup in the module |reals| has access to semantic macro |\inset| (membership) from the
module |sets| that was imported by the document by |\importmodule| directive from
the \url{../background/sets.tex}. Furthermore, it has access to the |\defeq|
(definitional equality) that was in turn imported by the module |sets|.

From this example we can already see an organizational advantage of \stex over {\LaTeX}:
we can define the (semantic) macros close to where the corresponding concepts are defined,
and we can (recursively) import mathematical modules. But the main advantage of markup in
\stex is that it can be transformed to XML via the {\latexml}
system~\cite{Miller:latexml}: Listing~\ref{lst:omdoc-ex} shows the
{\omdoc}~\cite{Kohlhase:OMDoc1.2} representation generated from the {\stex} sources in
listing~\ref{lst:stex-ex}.

\lstset{language=[1.3]OMDoc}
\begin{lstlisting}[label=lst:omdoc-ex,mathescape,escapeinside={\{\}},caption={An
XML Version of Listing~\ref{lst:stex-ex}},numbers=none]
<theory xml:id="reals">
 <imports from="../background/sets.{omdoc}#sets"/>
 <symbol xml:id="Reals"/>
 <notation>
   <prototype><OMS cd="reals" name="Reals"/></prototype>
    <rendering><m:mo>$\mathbb{R}$</m:mo></rendering>
 </notation>
  <symbol xml:id="greater"/><notation>$\ldots$</notation>
  <symbol xml:id="positiveReals"/><notation>$\ldots$</notation>
  <definition xml:id="posreals.def" for="positiveReals">
    <meta property="dc:title">Positive Real Numbers</meta>
    The set <OMOBJ><OMS cd="reals" name="postiveReals"/></OMOBJ> is the set $\ldots$
  </definition>
  $\ldots$
</theory>
\end{lstlisting}
One thing that stands out from the XML in this listing is that it incorporates all the
information from the \stex markup that was invisible in the PDF produced by formatting it
with {\TeX}.

\section{Tools for developing IDEs}

\subsection{Eclipse}
\index{Eclipse}
The Eclipse Project \cite{Eclipse:web} is a popular open-source
development platform integrating a set of extensible frameworks, tools
and runtimes for building, managing and deploying software. Based on
Eclipse development platform, many IDEs for different programming
languages exist today, most notably IDEs for Java and C++. The project
has a powerful plugin system conforming to the OSGi \cite{OSGi:web}
specifications which means that developed plugins can be used inside
other systems implementing the OSGi standard. Eclipse also includes
specialized tools for developing Eclipse plugins. These tools provide
support for importing functionality from other plugins, specifying
dependencies, assist at creating or extending menus or Eclipse
property pages as well as help to package plugins for easy web-based
installation. 

\subsection{Google Guice}
\index{Google Guice}
Google Guice \cite{GoogleGuice:web} is a dependency injection
framework for Java. It is heavily used by \xtext{} (see
\ref{relwork:sec:xText}) and also by \stexide{} to manage
inter-component dependencies as well as for creating mock objects
while testing. The main idea behind Google Guice to use Java
annotations to explicitly mark inter-component dependencies and let
the framework instantiate objects during run-time. For example, the
code in figure \ref{fig:relwork:inject} shows how dependencies are
made explicit by making them arguments of the constructor method and
adding the @Inject annotation to method definition. The last line,
shows how an instance of the {\tt SomeModule} object is created. The Google
Guice framework handles the creation of the {\tt Indexer} object and passes
it as first parameter to the constructor. The Google Guice framework
makes testing much easier because one can easily instruct the
framework to create a mock instance of the Indexer class when testing
the {\tt SomeModule} class. By that, we can make sure that our test
is only verifying the implementation of the {\tt SomeModule} and not
of its dependencies.

\begin{figure}
  \centering
\begin{lstlisting}[language=java]
Class SomeModule {
  Indexer indexer;

  @Inject
  public SomeModule(Indexer indexer) {
    this.indexer = indexer;
  }
}
...
SomeModule module = MyInjector.createObject(SomeModule.class)
\end{lstlisting}
\caption{Google Guice injection example to get an instance of the
  indexer object}
\label{fig:relwork:inject}
\end{figure}

\subsection{xText}
\index{xText}
\label{relwork:sec:xText}
\xtext{} \cite{xText:web} is a language development framework which makes
it easy to develop full-featured, Eclipse-based editing
environments for domain specific languages (DSLs). By domain 
specific language we mean a programming or specification language which
is dedicated to dealing with a particular problem domain. In
comparison to general purpose languages like Java or C++, the DSLs
design language grammar in such a way that domain concepts and
notations map naturally to the grammar symbols and rules. In this way,
when writing programs in the domain specific language, one directly
speaks of, for example, ``events'', ``guests'' or ``sessions''. Even
though DSL might be quite restrictive (depending on the domain and
application at hand), the only technical limitation imposed by
\xtext{} is that the language can be expressed by a Context Free Grammar
(CFG). 

One of the biggest achievements of the \xtext{} framework is that by simply
specifying an Extended Backus-Naur Form (EBNF) of the DSL grammar, one
can generate an editing environment which supports
many of the usual features like syntax highlighting, autocompletion 
as well as validation. The framework is highly configurable and allows
the developer to customize most of the behavior. Due to the fact that
\xtext{} relies heavily on Google Guice for dependency injection,
one can theoretically customize any component from \xtext{}
architecture. 

A typical \xtext{} project is split in two parts, each of them
represented by an Eclipse project. The first project, called DSL
project, generates an implementation of the DSL grammar which consists
of a language parser combined with classes which allow customizing
behavior of syntactic highlighting, autocompletion, validation
etc. The DSL project, is Eclipse framework independent and can be used in
any other stand-alone Java based program. The second Eclipse project,
imports the functionality from the DSL project and uses visual
components of the Eclipse environment to generate the editing
environment. 



\ifpdf
    \graphicspath{{X/figures/PNG/}{X/figures/PDF/}{X/figures/}}
\else
    \graphicspath{{X/figures/EPS/}{X/figures/}}
\fi




\chapter{System Implementation}

\section{Requirements}
\label{sec:sysimpl:req}
\index{Integrated Development Environment}
\index{sTeXIDE}
In section \ref{sec:aims:motiv} I discussed how previous experience
on authoring flexiform documents posed similar challenges to authoring
programs in object oriented languages like C++ or Java. Important similarities
include modularity and reuse of exiting code/knowledge, need to handle
large amount of interconnected components, need of a source building
mechanism. That lead to the idea of developing \stexide{} -- an IDE
for \stex{} similar to the ones for C++ or Java but with strong
emphasis towards extensibility and semantic services. By analyzing the
research questions from section \ref{sec:aims:aims}, a new list of
software requirements and features was compiled: 
\begin{compactenum}
  \item [\textbf{I1}] modular architecture tuned towards maximal independence of
    \stex{} language and portability for other editors. This
    requirement is only implicitly specified in the research aims by
    abstracting from \stex{} language and any programming platform. 
  \item [\textbf{I2}] create an extension mechanism allowing developers to extend
    IDE functionality with minimal knowledge about the internals of
    the IDE. This requirement is targeted towards minimizing the costs
    of developing new features and increasing the benefits of using
    the IDE.
  \item [\textbf{I3}] develop features which provide precise information about the
    context, depending on the edit position. 
  \item [\textbf{I4}] provide a mechanism for detecting redundancies and check for
    validity of \stex{} objects and their relationships. 
  \item [\textbf{I5}] enhance user experience by providing ways of fast navigation
    between objects. Visualize modules and dependencies by rendering
    theory graphs. 
  \item [\textbf{I6}] provide semantic search functionality
\end{compactenum}

\section{Implementation Approach}
\index{xText}
When designing the architecture of \stexide{}, we took special care to
make most of the architecture components independent from the
underlying language i.e. \stex{}. The main reason for it was that 
in the process of developing an IDE for \stex{}, we could also create a
language independent set of components which later could be reused for
other modular languages similar to \stex{} (e.g. \casl{} \cite{CaslTCS03}). Our
next priority was to provide an extension mechanism by which other
developers could contribute to the semantic services of \stexide{}
without going too deep into the IDE implementation. The value we see
in this extension mechanism for \stexide{} is that users can introduce
their own semantic macros and write small snippets of code and get IDE
support. 

The implementation of \stexide{} is based on the \xtext{} framework for 
Eclipse. The \xtext{} framework is made to be language independent by design
and hence provided a very good start for our own developments. We tried
to use the facilities provided by the \xtext{} framework at its most, 
hence winning more time for writing semantic services. Hence many
architecture components of \stexide{} are merely extensions of
\xtext{} components. Still, the extension mechanism, as well as indexing
support had to be implemented from scratch and represents a
considerable part of the contribution of the current work. 

In this section we will present in detail the main components of
the architecture and the decisions we took in order to achieve our
goals i.e. language independence and extensibility. In section
\ref{sec:arch:life}, we will introduce the components of the
architecture and on which user events they respond. The next 
subsections will discuss each of the components separately and will
describe their functionality and how they interact with the rest of
the components. 

\section{Architecture of \protect\stexide{}}
\label{sec:arch:life} \index{Abstract Syntax Tree}
The main components of the \stexide{} architecture are: 
\begin{compactenum}
  \item [\textbf{document parser}] parses \stex{} source and
    creates an Abstract Syntax Tree (AST).
  \item [\textbf{tagger}] assigns URI tags to AST nodes according
    to their semantic function.
  \item [\textbf{semantic syntax highlighter}] formats and colors \stex{}
    source code according to tags assigned by the tagger.
  \item [\textbf{validator}] performs consistency, redundancy checks
    and issues error/warning messages.
  \item [\textbf{context-aware autocompleter}] provides on-demand,
    context-sensitive suggestions on ways to complete \stex{} code near cursor.
  \item [\textbf{handler registry}] is the gateway between
    architecture components (tagger, validator, indexer etc) and user extensions.
  \item [\textbf{indexer}] indexes relevant parts and object
    relationships of \stex{} source. 
  \item [\textbf{semantic search}] provides advanced semantic
    searches in the corpus. 
  \item [\textbf{source builder}] responsible for converting \stex{} document
    in one of the output formats (.ps, .pdf, .omdoc, .xhtml etc)
\end{compactenum}

\begin{figure}[h]
\centering
\def\svgwidth{\columnwidth}
\input{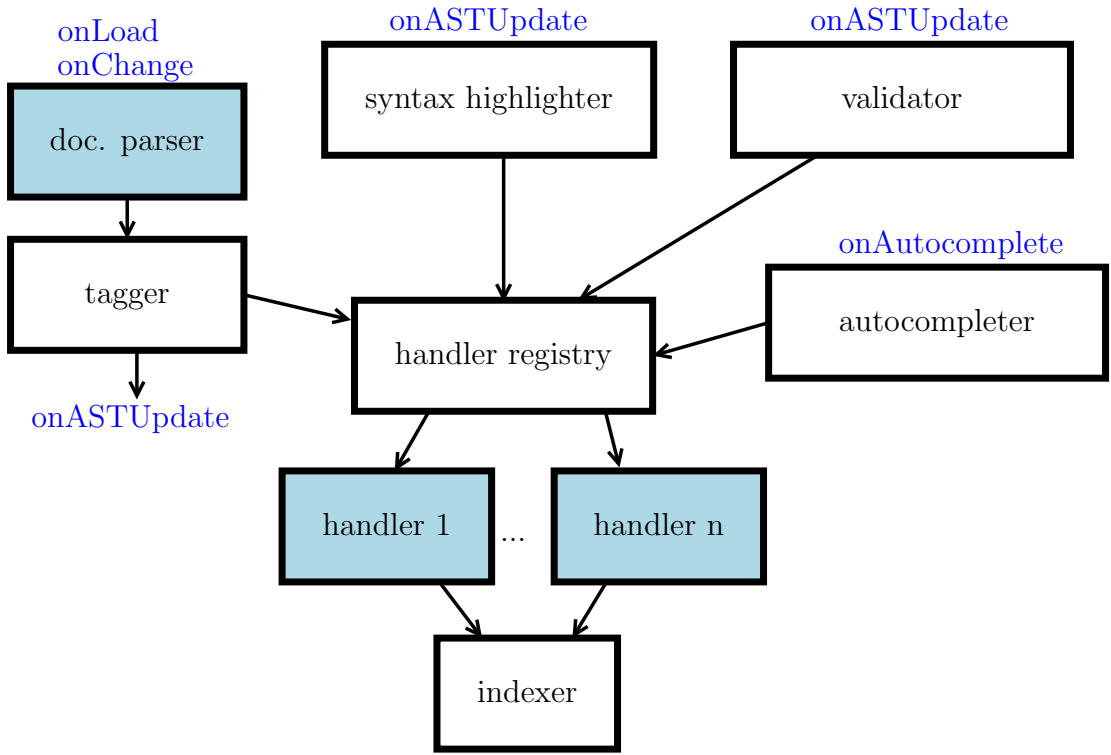}
\caption{Main architecture components and events to which they
  respond. Boxes colored in gray represent language dependent
  components which cannot be reused.}
\label{fig:arch:arch}
\end{figure}

In figure \ref{fig:arch:arch} one can see how different components
interact and form workflows. The workflows are triggered as a result
of user actions, namely: onLoad, onChange and onAutocomplete. The
onLoad event is created when a document is loaded into the IDE for the
first time; onChange gets triggered when user updated the document by 
inserting/deleting characters and finally onAutocomplete is created
when user requested autocompletion support. An additional internal event,
onASTUpdate, is triggered when the Abstract Syntax Tree corresponding
to a document is updated. 

As one can see from the figure \ref{fig:arch:arch}, the handler
registry and indexer components are part of each workflow and hence
central to the whole architecture. Also notice that in figure
\ref{fig:arch:arch}, we colored in gray the components which are
language dependent and hence have to be rewritten in case we want to
adapt the current architecture to a new modular language. For more
detailed information on what is the functionality and responsibilities
of each particular component, see next subsections.

\section{Document parser}
\label{sec:arch:doc_parser}
The document parser component is responsible for parsing and creating
the Abstract Syntax Tree (AST) of the input document. As all other
components of the architecture work only with the AST
representation, the AST has to be kept in sync with the input document
at all times. Hence the document parser component is the first one to
be executed after a document is loaded or modified. Several features
of the parser are extremely important, namely:

\begin{description}
\item [error recovery.] Most of the times, during the authoring process, the
  document contains errors which would make the parser reject the
  input string. However, that also means that no AST is created and so
  we cannot pass it further to other components. This is certainly
  undesirable since the IDE will not be able to assist the user in any
  further tasks. This is arguably the most important feature a parser
  for an IDE should have.
\item [efficient partial updates.] In the process of writing a
  document, changes happen frequently, however, most of them barely
  change the AST tree. This observation is an important optimization
  and one should definitely take advantage of it. 
\end{description}

As mentioned before, we based our implementation on the \xtext{} framework
which already comes with a mechanism of generating parser code and 
a set of classes which will later populate the Abstract Syntax Tree. 
The ANTLR generated parser is known to be very flexible in case of errors,
supports partial updates and has relatively good performance. Hence in
the case of the \stexide{} parser, we only had to specify the language grammar
and reused the parser functionality provided by the \xtext{} framework.

A full featured \LaTeX{} parser heavily depends on the context which can be
changed, for example, by importing other files. Let us examine the \LaTeX{}
code in listing \ref{fig:arch:latex_parser}. In the first line, the
'\%' symbol has the default meaning of begining of a comment. Assume that the
imported file 'redefine\_symbols' redefines the semantics of symbol
'\%' to actually print a '\%' character. Then the 3rd line will actually
be printed. Such context sensitiveness makes the parser extremely
powerful however also means that to parse a file correctly one also
has to parse all the dependencies. Such a parser can hardly achieve
the performance requirements needed for an IDE for keeping the source
and AST in sync. As \stex{} is just an extension of \LaTeX{} and the
ANTLR parser could be used only for CFG grammars, we decided to
develop a CFG  grammar which approximates at its best the common
practices of writing \stex{} documents. This grammar essentially
parses \TeX{} commands, their options as well as text and puts them in
a AST.

\begin{lstlisting}[language=sTeX, caption={Example of context
sensitiveness of \LaTeX}, label=fig:arch:latex_parser ]
The following input command redefines the percent symbol 
% here \% means a comment
\input{redefine_symbols}
% this is not a comment any longer 
\end{lstlisting}

The approximated \stex{} CFG grammar is relatively simple. Namely, there
are 4 types of objects: Model, Word, Command and Option. A simplified
version of the grammar is given in listing \ref{fig:arch:stex-grammar}.
The grammar used in the implementation removes the ambiguities of the
simplified grammar as well as tackles issues like \unicode{} symbols, 
comments, expressions like '\textbackslash['etc. As one can see, our
grammar does not include rules for matching \textbackslash
begin\{envname\}, ... \textbackslash end\{envname\} statements. The
main reason for it is that environment mismatches are generally hard to resolve and
a generic parser algorithm would not be able to handle them
graciously. Our solution was to keep the grammar simple (hence more
error tolerant) and implement a specialized algorithm to match 
maximally many \textbackslash
begin\{envname\}, ... \textbackslash end\{envname\} pairs and provide
meaningful error messages in case of mismatches. 

\begin{lstlisting}[language=java,numbers=none, caption={Simplified
version of the approximated \protect\stex{} CFG grammar}, label=fig:arch:stex-grammar]
Model = (Word | Command)*
Command = \Word Option*
Option = { Model } | [ Model ]
\end{lstlisting}

The decision to make \stex{} CFG grammar as simple as possible proved
to be a good design choice as it shifted handling of macros towards the 'handler
registry' plug-in mechanism which we describe in the following section.

\section{Handler Registry}
\label{sec:arch:handler-registry}
A very important requirement for the architecture of \stexide{} was to
create a mechanism by which users can extend editor's
functionality. The handler registry component represents the core of
the extension mechanism and is responsible for loading \stexide{}
extensions, creating a catalog of what AST nodes each extension is
responsible for as well as provide instances of extension objects. 

The Handler Registry component represents the gateway through which 
all other software components (tagger, syntax highlighter, validator etc) 
can pass control to specialized handlers to customize behavior.
Consider the example when the syntax highlighter component encounters an AST
node defining a new \stex{} module. The component requests the Handler
Registry for the extension handling that AST node and gets an instance
of the ``module definition'' extension. This instance, is responsible
for specifying any custom behavior for the AST node and hence the
highlighting component will request it to specify what colors to use
for highlighting of module definitions. Likewise, when the validator
component encounters the same AST node, it will get an instance of the
same extension and will ask it to check if the AST node contains 
semantically correct information (e.g. if imported file exists).
By that, all the custom behavior is shifted to the plugin modules
and hence makes \stexide{} very extensible and many of the
architecture components reusable. 

The Handler Registry implementation stores a list of objects 
implementing the IExtension interface. This is the interface
\stexide{} plugins should implement. The main methods to be
implemented are summarized in listing \ref{fig:arch:iext}. Each
of the components call one or several of the interface methods
to customize their behavior. For example, the syntax highlighter
calls the getSyntaxColorURI method, while the validator calls
the validate method. 

\begin{lstlisting}[language=java, caption={Relevant parts of the IExtension interface}, label=fig:arch:iext]
public interface IExtension {
  String[] getHandledCommandNames();
  String[] getHandledTags();
  String[] getHighlightingURIs();

  void addNodeTags(Command cmd);
  String getSyntaxColorURI(String tag);
  Boolean index(String tag, IPropertiesAcceptor properties);
  void validate(String tag, EObject object, AbstractLaTeXJavaValidator validator);
	
  void autocompleteTag(String tag, AbstractNode lastComplete, 
                       String prefix, IAutocompletionAcceptor acceptor);
	
  void refactor(String tag, EObject obj, 
                IDialogProvider dialogProvider);
}
\end{lstlisting}

\section{Tagger}
\label{sec:arch:tagger}
After parsing the document, we get an Abstract Syntax Tree which contains
only basic structural information about the document. This structure
knows only about words, \stex{} macros ({\verb Command } objects) and
their options. The real semantics behind those commands are still
undefined and it is the responsibility of the Tagger component to
assign semantics to AST nodes. In our case, semantics is given by
assigning AST nodes some tags (URIs) which specify the semantics of
the nodes. As mentioned in section \ref{sec:arch:handler-registry},
the handler registry component should be able to identify the plugin
to be used for customizing the behavior at certain AST nodes. This is
achieved by using the tags assigned by the tagger component.

In \stexide{} we made the assumption that the
most interesting information is provided at or near certain
commands. That is why, each plugin first specifies which commands are
``interesting'' for it and then gets called by the tagger component to
add extension specific tags to the AST tree nodes. That is done traversing the AST 
generated by the parser, identifying the commands which are interesting for 
certain plugin and executing the  {\tt addNodeTags} method of that plugin on 
each of those commands. As the name of the method already suggests, the 
plugin is afterward responsible for tagging the really useful information
which might be certain option parameters etc. 

Let us assume that we want to implement a plugin responsible for managing
\stex{} definitions (see listing \ref{fig:arch:ex_definition}). The \stex{} 
command might specify several important facts about the definition. First, the
concept for which this definition holds (the 'for' key-value pair in the command
 options) as well as the text of the definition. Then, such a plugin should
return 'definition' as one of the { \tt getHandledCommandNames}
results. The {\tt addNodeTags}
method should tag the value of the 'for' key-pair with a URI 
(ex. {\tt stex.definition.definitionfor}) and the whole content
with a tag e.g. {\tt stex.definition.definitionText}. 
In the same time, the plugin should return both of these tags in the
result of the
{\tt getHandledTags} method so that hander registry knows to pass control to the 
plugin when encountering those tags. 

\begin{lstlisting}[language=stex, caption={Sample \protect\stex{}
definition}, label=fig:arch:ex_definition]
\begin{definition}[id=functions.def, for=fun]
  A {\defin{function}} $\fun{f}AB$ is a left-total, right-unique relation in $\cart{A,B}$
\end{definition}
\end{lstlisting}

\section{Semantic Syntax Highlighting}
The implementation of the semantic syntax highlighter extends the
highlighting mechanism of \xtext{}. The extension implements
a method which assigns to each AST node, a certain semantic
category such that text parts belonging to the same category are to be
colored with the same color. It is the duty of \xtext{} to further
take care that the text of certain category gets colored and
styled the way the user specified it in the IDE preferences dialog. 

\begin{figure}
\centering
\includegraphics[width=10cm]{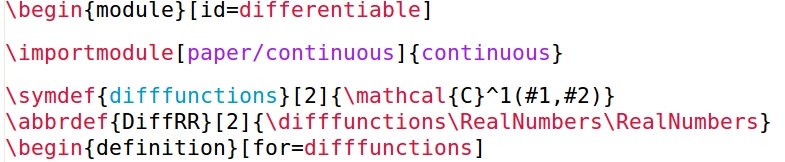}
\caption{Semantically highlighted \protect\stex{} source}
\label{fig:arch:syntax}
\end{figure}

The semantic syntax highlighter component is run every time the
AST tree is updated. At the time when the component is executed, the 
AST is already ``enriched'' with tags. The main task of the semantic
syntax highlighter component is then to map tags of the AST nodes to 
one of the category URIs known by \xtext{}. Since tags assigned to AST
nodes have semantic meaning, then by merely mapping them to colors
gives us semantic syntax highlighting (see figure
\ref{fig:arch:syntax}). The mapping is done as follows: the AST is 
traversed and for each node having a tag we use the Handler Registry
to get the handler responsible for providing the category URI.
The method from IExtension interface responsible for semantic syntax
highlighting is {\tt getSyntaxColorURI}. It get as parameter the tag
of the current AST node and returns the category URI. 

\section{Validator}
\begin{figure}[h]
\centering
\includegraphics[width=10cm]{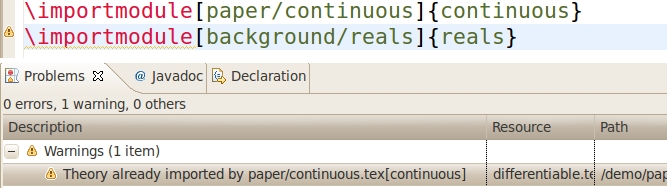}
\caption{Validation feature recognizes a redundant import}
\label{fig:arch:validator}
\end{figure}

The validator component is responsible for performing sanity checks of
the \stex{} input without running a full featured \LaTeX{}
process. The checks are aimed at providing immediate error/warning
messages for typical sources of errors. The checks can become
arbitrarily complex and easily surpass the analysis capabilities of
\LaTeX{} workflow. For example, in \stexide{} we implemented a
validator which checked for redundant module imports and issued
warnings at importmodule statements which were redundant (see figure
\ref{fig:arch:validator}). This type of warnings are currently not
being issued by the \stex{} extension and 
would probably take too much effort to implement in
\LaTeX{}. Another advantage of validator extension over error/warning messages
parsed from the output of \LaTeX{} is that they usually provide better
explanation of the error as well as how to correct it. 

From an implementation point of view, the validator component
resembles a lot the semantic syntax highlighting component. It
extends the validator implementation of \xtext{}, traverses the
AST, uses Handler Registry to get hold of the appropriate handler and
gives it a way to add warning and error messages. The method from the
IExtension interface responsible for validation is {\tt validate}.  It
takes 3 parameters, namely: the tag of the AST node, the AST node
itself (so that \xtext{} can compute the position where to display the
error/warning message) as well as a class providing functionality for
adding new errors/warnings.

\section{Context-sensitive autocompletion}
The next, useful and central feature of \stexide{} architecture is
the context-aware autocompletion feature. The most useful bit is that
autocompletion is context-aware i.e. it understands what kind of
content is expected at certain positions in the \stex{} document and
exactly that type of content is suggested. For example, consider the
\stex{} input in listing \ref{fig:arch:autocomplete}. If one tried to
autocomplete at line 2 of the input, \stexide{} would suggest just
several context-independent \LaTeX{} macros. If one would try to
autocomplete while in the first argument of the {\tt \tbs importmodule}
command, the suggestions will be paths to valid files on the file
system. On the other hand, the second argument of the same {\tt \tbs
importmodule} command would only suggest valid modules from the file
specified in the first argument. Likewise, if one would try to
autocomplete at the end of line 3, the autocomplete feature 
will suggest semantic macros like \tbs in or \tbs union which were
imported from the 'sets' module. As one can see, the suggestions of
the autocomplete feature might differ a lot within 1 line depending on
the context. To authors knowledge, no other IDE for \LaTeX{} supports 
context based autocompletion at this level of granularity.

  \begin{lstlisting}[escapechar=|,language=sTeX, caption={An
\protect\stex{} module for Real Numbers}, label=fig:arch:autocomplete,
numbers=none]
\begin{module}[id=reals]

  \importmodule[../background/sets]{sets}
  \symdef{Reals}{\mathcal{R}}
  \symdef{positiveReals}{\Reals^+}
  \symdef{greater}[2]{#1>#2}
  \begin{definition}[id=posreals.def,title=Positive Real Numbers, for=positiveReals]
    The set $\positiveReals$ is the set of $\inset{x}\Reals$ such that $\greater{x}0$
  \end{definition}
  |\ldots|
\end{module}
\end{lstlisting}

\begin{figure}[h]
\centering
\includegraphics[width=12cm]{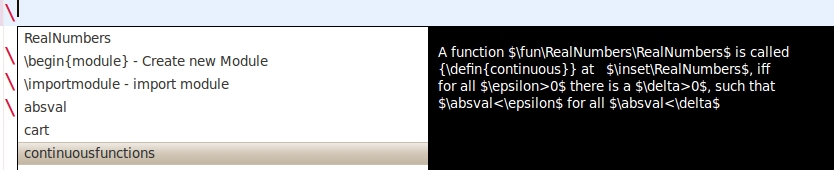}
\caption{Autocomplete feature showing definition of continuous
  functions next to the \protect\stex{} macro}
\label{fig:arch:autocomplete_def}
\end{figure}

Beside assisting user with context sensitive autocomplete suggestions,
the module also performs Information Retrieval tasks. Namely, when a
certain \stex{} macro is suggested during autocompletion, it also
searches for definitions which explain that macro. This information is
shown to the user when he/she has the cursor over the macro name (see
figure \ref{fig:arch:autocomplete_def}). This feature is very useful
when unsure of the real semantics of the \stex{} macro, which happens
a lot to beginners but is also useful for power-users.

The context sensitive autocompletion is also implemented as an
extension of the \xtext{} framework. The \xtext{} framework provides
methods which are executed when autocompletion is requested by the
user and provides information about the context where the user
requested autocompletion. Namely, one can get the position of the
cursor as well as the AST leaf node preceding the location where
autocompletion was requested. Having this information, the \stexide{}
autocompletion component analyzes the AST leaf node given by
\xtext{}, gets the handlers responsible for tags assigned to the node,
runs them and merges their results. The method in the IExtension
interface responsible for autocompletion is {\tt autocompleteTag}. It
gets as parameters the tag to be autocompleted, the leaf AST node as
well as the prefix autocomplete suggestions must have. The last
component is an interface which accepts suggestions and optionally
also explanatory text to be shown when user selects the suggestion.

\section{Indexer}
\subsection{Motivation}
Many of the features currently supported by \stexide{} are only
possible (or make sense) in the context of a large document corpus. For example, 
the validator needs to check if the module to be imported actually
exists in the specified file; or autocompleter needs to know what concepts
were defined in imported modules so that it can display them for
autocompletion. In order for these features to function, they need to
access data from other files. The easiest way to do that, is to 
parse the necessary files and search for the relevant information in
the parsed AST trees. That would, in fact, mean to run the document
parser as well as the tagger. Suppose this approach were used for
context sensitive autocompletion and we had the task of autocompleting an
\stex{} macro. To do it properly, one would have had to find all concepts in
the current module defined before the cursor and merge them with concepts
from imported modules. Now, if imported modules, imported other
modules and so on, we would get to the situation where he had to
parse 50\% of the corpus. Even though the document parser and tagger
components are relatively fast, doing so for tens of files has an
obvious performance impact. In the \gencs{} corpus, such an
unoptimized behavior takes about 3-4 seconds. 

To optimize these processes, I introduced an index component which is
responsible for giving fast and targeted access to data in
documents. The set of requirements for this component were:
declarative way of querying data, performance and scalability, but the
most important, give total freedom to \stexide{} extensions to choose
what data to index as well as provide an easy interface for inserting the
data. 

\subsection{Design decisions}
The first design decision to be made was to choose the kind of
structures we wanted to index. Two alternatives we considered, namely:
storing tree structures 
resembling parts of the AST trees or storing a relational table
specifying the type of the information stored as well as its value and
location. The first approach has the advantage of keeping the
structure of the indexed information close to the structure of the AST
and hence document. That makes querying more intuitive as well as
powerful. For example, the linear structure approach would hardly be
able to answer a query like ``give me the IDs of all concepts defined
in module X from file F'' because the query would have to contain
constraints (ex. comparison of location) to find out if a
concept is defined inside module X and not any other module defined in
file F. The relational table approach, however, would have much better
performance, would require a much simpler interface for inserting data
and is easier to implement. After considering both alternatives we
chose to store tree structures because querying the index using the 
linear relational table approach would have been too difficult and
error-prone. We considered performance to be of secondary priority for
now and decided to direct our effort into creating a simple to
understand interface for extensions to store their data. 

After the first design decision was made, we had to choose how we wanted
to store and index tree structures? Two options we considered 
were using a RDF or XML database. Both provided a declarative query
language (SPARQL and XQuery), could index trees and were flexible about
the structure and type of information to be indexed. The advantage of
using an XML database is that tree data structure is the natural unit
the database understands, whereas in RDF databases it is the ``subject
predicate object'' triple. On the other hand, in a RDF database, many
of the relations between files could be made explicit. For example,
consider module A imports module B and uses a certain concept X from
B. Then, every time concept X were used in A (call that instance Y), an
additional triple could be made saying ``Y isSameAs X''. So if we ever
wanted to know where a certain concept X was referenced, we only had
to query for relationship ``? isSameAs X''. Also, an earlier attempt
of the author to index trees by storing them in an XML database showed
that in some cases querying them using XQuery was quite intricate. So
the final decision was made to use an in-memory RDF database based on
Jena \cite{URL:jena:web}, called TDB.

\subsection{Challenges}
An important challenge we had to solve was how to orchestrate the
creation of the tree to be indexed. The problem is that we have
several independent handlers which want to control how their data is
indexed. On the other hand, they still have to create a tree
structure similar to the one in AST. So the question is how much
control should be given to handlers? Part of the solution was to limit
the control of handlers over tree creation by only letting them specify whether
the node they are responsible for, should be indexed or not. In this
way, the indexing component is able to create a similar tree
structure as the original AST and just omit nodes which should not be
indexed. This removes the responsibility from the handlers to keep the
right tree structure as well as hides the RDF inability to naturally
represent trees.  

\begin{lstlisting}[escapechar=|,language=sTeX, caption={An
\protect\stex{} module for Real Numbers}, label=fig:arch:definiendum, numbers=none]
\begin{module}[id=sets-operations]
  \symdef{cart}{\times}
  \begin{definition}[id=Cartesianproduct.def,display=flow,for=cart]
    {\twindef{Cartesian}{product}:}
    $\defeq{\cart{A,B}}{\setst{\tup{a,b}}{\conj{\inset{a}{A},\inset{b}{B}}}}$, call
    $\tup{a,b}$ {\defin{pair}}.
  \end{definition}
\end{module}
\end{lstlisting}

\begin{figure}
  \centering
  \includegraphics[width=\textwidth]{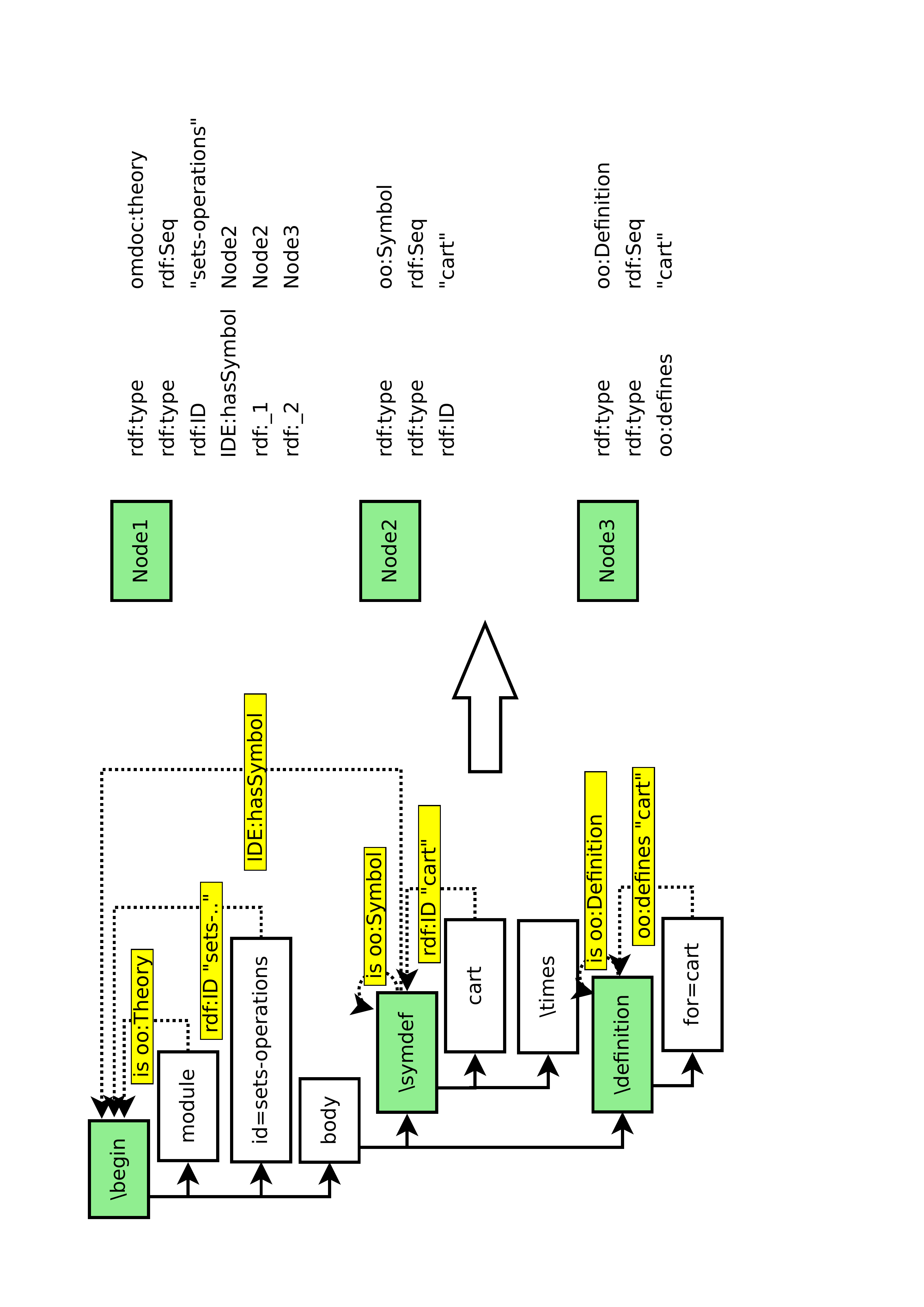}
  \caption{Index tree RDF representation. }
  \label{fig:arch:indextree}
\end{figure}

The problem with this approach is that usually the information we want
to index is ``spread around'' several AST nodes even though they
represent information about one concept. Consider the \stex{} code in
listing \ref{fig:arch:definiendum}. The root of the AST tree (see
figure  \ref{fig:arch:indextree}) is a
{\verb Command } node with 3 {\verb Option } children (the 3rd 
{\verb Option } being the body of the module until \tbs
end\{module\}). The part of text covered solely by the root AST node
is {\tt \tbs begin} which obviously does not give much information. The
important bits of information needed to make module definition
complete should come from the 1st and 2nd {\verb Option }
children. The only way to index the information with aforementioned
approach would be by creating an RDF node for both of them. This would
increase the complexity of the RDF tree, would make it more confusing 
and more difficult to query. A  solution to this problem would be to
choose only one of the AST nodes defining a concept as ``responsible''
and let it search the right information in the AST tree. In our
example, we could choose the root AST node ({\tt \tbs begin}) as
responsible and let it look at the first and second children arguments
for information. Furthermore, due to the fact that all modules tag parts
of text with unique URIs, we could also make links to information from
other AST nodes.  For example, for module definitions it is useful to
index  what symbols are defined in it because this information is
often used for autocompletion as well as validation. In our case, we
could make a link from the module definition RDF node ({\tt sets-operations}) to
the symbol definition RDF node ({\tt cart}) with predicate relation
{\tt oo:partOf}.

A big drawback of this approach is that the responsible node has to
know about other node types and what URIs they use to denote their
data. That breaks module independence principles because as soon as a 
module changes its URIs, the code inside the responsible node will
fail. It also breaks extensibility principles because new modules, which
want to link with some responsible AST node, will have to update bits
of code in the implementation of the responsible AST node. To solve
all these issues we decided to make the non-responsible AST nodes to
search for the responsible ones and communicate to them additional
information to be indexed. In this way, responsible AST nodes only
have to provide a general mechanism for adding new properties as well
as linking with other RDF nodes; and non-responsible nodes, who have
control over their data, can add any useful data they wish to any of
the responsible nodes. By that, extension handlers are relatively
unlimited in the type of information they get indexed, AST structure
is maintained and modularity and extensibility of the extensions is
preserved. 

To get a better intuition how the generated RDF looks like, let us
examine figure \ref{fig:arch:indextree} in more detail. On the left
side of the figure one can see a fragment of the parsed AST
corresponding to \stex{} source in listing
\ref{fig:arch:definiendum}. The green boxes represent the AST nodes
for which the assigned extension handler specified that the node should be
indexed. Indeed, the RDF representation of the AST on the right side
of the figure contains 3 RDF nodes corresponding to each green
AST node. The solid arrows in the AST tree represent the parent-child
relations in the original AST tree. In the RDF representation, the
AST structure is kept by adding to each RDF node the type
{\tt rdf:Seq} which denotes that the node contains an ordered sequence
of objects. The first object in the sequence is specified by the
predicate {\tt rdf:\_1}, the second object by the predicate {\tt rdf:\_2}
and so on.  We use this sequence to specify the children of the each
of the RDF nodes. In our example, the resulting indexed tree is quite
simple, namely, it just has a root node ({\tt Node1}) with 2 children
({\tt Node2} and {\tt Node3}). In the original tree, the AST nodes
corresponding to {\tt Node2} and {\tt Node3} are, however, not direct
children of {\tt Node1}. This is due to the fact that the {\tt body}
AST node was not indexed and hence {\tt Node2} and {\tt Node3} were
attached to the nearest ancestor which was indexed. This also implies
that possible siblings in resulting RDF 
tree are not necessarily siblings in the original AST tree. So the
question is ``what did we preserve? what was the value of keeping the
AST structure?\protect``. Indeed, the AST structure is only partially
preserved but in fact, since only important parts of the AST are
indexed, the RDF tree captures the core relations. If we look at our
example, the RDF tree reads: ``we have a theory object with two
children. First one defines a symbol 'cart' and the second is a
definition''. This is much better then having to say ``the theory
object has a body, which in turn has 2 children'' because the later
captures a lot of useless technical intricacies. 

Another interesting question is: how stable is the indexed RDF tree
structure? can one rely on it for querying? The answer is
unfortunately: no, one cannot rely on the tree structure for
querying. The reason is simple, as soon as a certain AST node is
considered important enough to be indexed (ex. by adding a new
extension to the Handler Registry), the tree structure changes. For
example, in figure \ref{fig:arch:indextree}, the last child of the
root node labeled {\tt body} is not indexed, hence the {\tt \tbs symdef} and
{\tt \tbs definition} nodes connect directly to the {\tt \tbs begin node}. If,
however, a new extension is loaded which marks the {\tt body} node as
important, then both {\tt \tbs symdef} and {\tt \tbs definition} node will have to
connect to the {\tt body} node. Hence the tree structure is changed.
One solution to solve this problem is to always keep this limitation
in mind while querying and instead of querying for ``give me the child
of type X'' one could query ``give me a descendant of type
X''. Unfortunately the later is very inefficient to implement. The 
second solution, the one we use in \stexide{}, relies on semantic
inter-node connections like ``Node1 IDE:hasSymbol Node2'' from our
example. This relation was generated at the point when the {\tt \tbs
symdef} AST node was indexed. It searched for the first ancestor
defining a module and connected to it by adding the {\tt IDE:hasSymbol}
relation to it. The AST node handler did so because it expects this relation
to be useful when querying the index. At this point we
realize how important was the decision to make not indexed nodes to
search for the indexed ones and add information to 
them. In our case, the same mechanism was reused to also connect
indexed nodes in a reliable manner making queries stable towards
changes in the set of extensions. 

\subsection{Implementation}

From the implementation point of view, the indexer component of
\stexide{} was implemented in a similar way as all the other
components, namely, it provides only some basic functionality and
delegates control to extension handlers to specify what information
should be indexed. The method of the IExtension interface implemented
by all handlers inside Handler Registry component is called
{\tt index}. The first parameter is, as usual, the tag to be indexed and 
the second is an object implementing {\tt IPropertiesAcceptor} interface
which is used to communicate what information should be stored in the
index (see listing \ref{fig:arch:IPropAcc}). Unlike other methods, the
{\tt index} method also returns a boolean value which controls if a new
RDF node should be created for that particular AST node. 

\begin{lstlisting}[language=java, caption={The IPropertiesAcceptor
interface}, label=fig:arch:IPropAcc]
public interface IPropertiesAcceptor {
   void addIntegerProperty(String property_id, int value);
   void addStringProperty(String property_id, String value);
   void addLinkProperty(String property_id, IPropertiesAcceptor resource);
   void addResourceProperty(String property_id, String URI);
	
   public List<? extends IPropertiesAcceptor> getStack();
   public EObject getASTNode();
}
\end{lstlisting}

As one can see, the IPropertiesAcceptor interface gives possibility to
add different types of information. It can add constant properties
(integer \& string types), link to other resources identified by their
IPropertiesAcceptor object (used for making the connection {\tt Node1
IDE:hasSymbol Node2}) as well as a general method for adding any URI property. 
The {\tt getStack} method returns all the IPropertiesAcceptor objects
of parents of the current AST node which returned true at
their {\tt index} method. And the last method, {\tt getASTNode}, returns the AST
node identifying the current IPropertiesAcceptor object.

\section{Semantic Search}
\begin{figure}
\centering
\includegraphics[width=12cm]{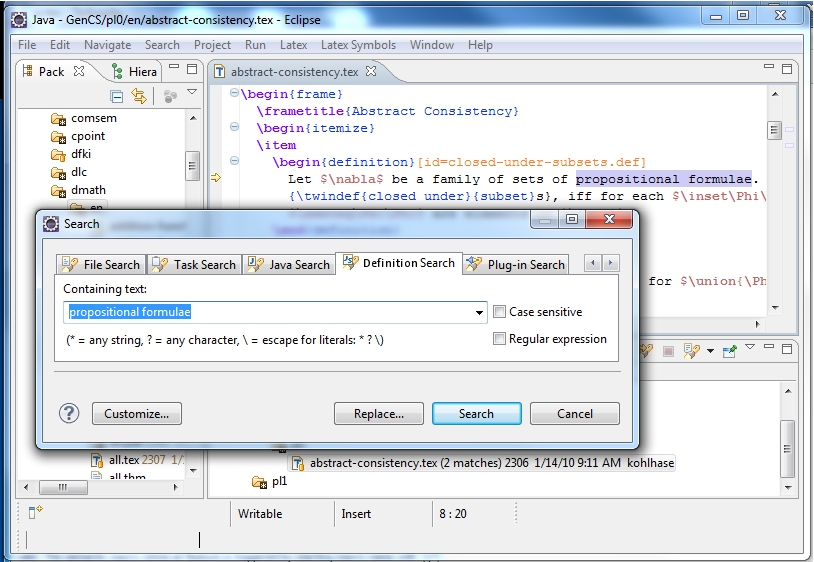}
\caption{Definition search mechanism of \stexide{}}
\label{fig:arch:defsearch}
\end{figure}

In this section I would like to present a mostly user interface
feature of \stexide{} which uses the indexer component to perform
semantic search. The idea is to provide ways to filter the type of content
that should be searched. The current implementation is merely a proof
of concept showing how one can perform definition search (see figure
\ref{fig:arch:defsearch}). 

From the implementation point of view it first makes sure that the
project index is up-to-date i.e. asks the indexer component to index
any changes in the documents. After this is done, it queries for all
the definitions and performs a simple key-word search.

\section{Build System}
An important feature of \stex{} is that one can generate a wide
spectrum of output formats. For example, one can use a typical
\LaTeX{} compiler like {\tt pdflatex} to produce presentation
oriented outputs like .pdf or .ps. On the other hand, by using
{\tt \latexml{}} and {\tt \latexmlpost{}}, one can generate XHTML documents
suitable for rendering in the browser and \omdoc{} documents -- for
exchanging knowledge between applications. In \stexide{}, I found it
important to give user the possibility of creating any of the
aforementioned formats with minimal efforts. In fact, whenever
possible, the user should be able to see the latest version of the
document in the chosen output format. 

To perform a on-the-fly conversion is generally not possible because
of the large number of files to be read and processed. This is,
however, a typical problem and a common solution other IDEs use is to
start the building process when the user stops typing (during idle
time). This means that the user does not get instant access to
the generated content but this is not considered as being a problem. 

The process of generating different output formats from \stex{}
sources can be seen as a sequence of program executions. Depending on
requested format and document properties, there might be  
different execution sequences for generating the same output format. For
example, to generate a .pdf file one can just run the {\tt pdflatex} program
on the \TeX{} source file. On the other hand, if the \TeX{} file
contains bibliography entries, it is recommended to run {\tt bibtex}
first followed by running {\tt pdflatex}. To generate XHTML or \omdoc{}
formats one uses the same sequence of programs ({\tt \latexml{}}, {\tt
  \latexmlpost{}})
with slightly different command line parameters. For \omdoc{} one
would additionally apply an XSLT style-sheet to the output of
{\tt \latexmlpost{}}. 

As one can see, there are many compilation workflows which differ by
the set of programs, command line parameters or even sequence in which
they are executed. On the other hand, tasks like parsing compiler
output/error messages of the programs remains unchanged and
independent on the workflow or parameters they get. So it is natural
to separate the code for program execution as well as output parsing
(called program handlers) from the general workflow mechanism. 

The workflow mechanism, apparently, has a very simple task, namely,
run code written in program handlers in a certain sequence. This is
however, only the tip of the iceberg. One of the features important
to be supported was to make it possible to provide input data through
a stream preferably without creating a temporary file on the file
system. In general, most of the tools we specified earlier are capable
of reading the input data from standard input. Some of them
(ex. {\tt pdflatex}) can only write output data to a file because it
actually generates several files. So each of the compilers
has its own limitations about supporting reading or writing from/to
standard input/output. Suppose A and B are two consecutive steps in a
workflow. Step A supports writing to standard output but the one from B
-- can only read from file. Then, during workflow execution, one would
have to redirect the output from A to a temporary file and provide the
name of the file to step B. In the builder component of \stexide{},
the workflow manager was implemented in a way that it can query individual
compiler steps for metadata which includes information on type of
input/output they support. Then, the component builds a execution plan
which takes care of passing correctly the output of one step to the
input of the next. 



\chapter{Discussion} 
In this chapter, I will try to estimate the costs of developing new 
extensions for \stexide{} from the user perspective. Namely, in
section \ref{sec:arch:user_perspective}, I will discuss the process of
implementing an extension and will analyze the effort a developer should
invest for each step. In section \ref{sec:discuss:conclusion}, I will
summarize the result of the current research by analyzing how the \stexide{}
system implementation could answer the research questions posed in
section \ref{sec:aims:aims}. I will conclude this thesis by a section
on future work.

\section{Extension mechanism: User perspective}
\label{sec:arch:user_perspective}
In the previous sections we described the extension mechanism we
implemented in \stexide{} and how different components use this
mechanism to customize their behavior. In this section, I would like
to comment on the usability of the extension mechanism i.e. to present
how easy or difficult it is to achieve some goals, as well as show
some best practices developers of new \stexide{} extensions should follow. 

As mentioned in previous sections, an extension should implement the
IExtension interface. Suppose we want to create an extension for
handling {\tt \tbs importmodule} commands. First step, is to implement the
{\tt getHandledCommandNames } method. This is extremely easy, as one
should just return the name of the command we want to handle, in our
case, 'importmodule'. Next, is the {\tt getHandledTags } method
which should return an array of unique URIs which we use for tagging different
parts of its data. The {\tt \tbs importmodule} command has 2 arguments,
namely, the path of the file containing the module to be imported and
the id of the module to be imported. We will need 3 Tags: one for
denoting the importmodule command in general (and will tag the {\tt \tbs
importmodule} part of the text), one for the file path and one for the
module id. For sake of simplicity lets call these tags {\tt importTag},
{\tt fileTag} and {\tt idTag} respectively. So the {\tt getHandledTags}
should return an array with these 3 tag names. The complexity of the
method implementation is minimal. The {\verb getHighlightingURIs }
method should return the URIs of the color category we want to use for
coloring the tags. We will use a default ``command'' category for coloring
the {\tt \tbs importmodule} part but want to introduce a new category
for both file path and module id options to denote external
references. So the {\verb getHighlightingURIs } will return 2 URIs,
one of which is not known to the system, yet. A text description for
the unknown color category URI will be given by the {\verb getHighlightingURIDescription }
method and so the user will be able to specify/change the desired
color for this category in the eclipse preferences
menu. Implementation complexity for both methods is again -- minimal. 

Next method to be implemented is the {\verb addNodeTags } method. The
method takes as argument a {\verb Command } object, which in fact
extends a normal AST node with methods for getting the command name,
as well as getting the options of that command. The {\verb addNodeTags }
should, in this case, use several utility methods provided in the
\stexide{} framework, to add {\tt importTag} to the {\verb Command }
object, {\tt fileTag} to the objects inside the first option and
{\tt idTag} to objects in the second option. Obviously one should take
precautions against some of the options missing but otherwise the
implementation is straightforward. Compared to the other 2 methods
we described until now, this method is a bit more complicated but
generally it is easy to implement it. 

Finally we come to the first method which will affect the user
interface, namely, the semantic syntax highlighting. The only thing
the method should provide is a mapping from tags given by the 
{\verb addNodeTags } method to one of the category color URIs returned
by the {\verb getHighlightingURIs }. This is straightforward and in
the same time gives developer the opportunity to ``test-drive'' the
implementation to check for inconsistencies. A very nice feature of
the developed framework is that to achieve semantic syntax
highlighting for the developed extension, one should just write some 60
lines of code (see appendix section \ref{minimal_syntaxhighlighting})
which are also easy to understand. Until now, the order 
in which the methods were implemented had to follow a certain logical
sequence. The rest of the methods in the IExtension are independent
from one another and should be implemented according to developer's
needs and priorities.  

The next comes the autocompletion feature. Since there are 3 tags for
which the {\tt \tbs importmodule} handler is responsible, we have the
possibility to create 3 different workflows to handle each case
separately. The first tag is {\tt importTag}. If we 
are asked to autocomplete it, this means that the text {\tt \tbs
importmodule} is already written and we asked for autocompletion
when the cursor is in the end of the {\tt \tbs importmodule} text. The
only way we can autocomplete it is by suggesting to open square
brackets and write a file name. So we create the list of top level
files and directories in the project, put square brackets around them
and return them as the list of autocompletion possibilities. Next
case, is if we are asked to autocomplete an object tagged with
{\tt fileTag}. This means that we are in an option object which contains
a list of words (forming project a relative path), where the last word
where the cursor is sitting is given by lastComplete object. We have
to concatenate the string values until the lastComplete node, create a
list of files in that directory and filter out those which don't match
the value in lastComplete. The last tag we have to autocomplete is the
{\tt idTag}. To be able to autocomplete it, we first have to find out
what is the file from which the module is imported. This is easily
achieved buy taking the contents of the first option node of the
parent. Let the file be {\tt A.tex}. Then, we have to look up for module IDs
in {\tt A.tex}. This is done by using the index and first we get hold of the
Indexer component object by declaring a Indexer object as part of the
constructor of the extension object and adding the @Inject annotation
(see figure \ref{fig:arch:indexerObj}). The Google Juice architecture
will make sure to pass the singleton instance of the Indexer factory. 

\begin{lstlisting}[escapechar=|,language=java, caption={Getting an
instance of the Indexer object.  }, label=fig:arch:indexerObj]
Indexer indexer;

@Inject
ImportModuleCommand(Indexer indexerObject) {
   this.indexer = indexerObject;
}
\end{lstlisting}
Then, we request the Indexer factory to get the index of the
{\tt A.tex file}. The indexer will check if it has the newest version of the
file already indexed. If it does not have it indexed, or it is outdated
then it parses the file, runs the tagger and then run the indexer on the
AST. In both cases, the indexer will return the RDF node coresponding
to the up-to-date AST root of the {\tt A.tex} file. Then we just have to run
a SPARQL query which one can see in listing
{\ref{fig:arch:indexerSparql}}:

\begin{lstlisting}[escapechar=|,language=java,caption={SPARQL query getting the IDs of modules defined in a document
  with root AST node coresponding to indexed ASTRootURI RDF node.}, label=fig:arch:indexerSparql]
SELECT ?moduleId WHERE 
          <ASTRootURI>        IDE:hasModule           ?y
	   ?y                 rdf:type                oo:Theory
	   ?y                 rdf:id                  ?moduleId
\end{lstlisting}
The persieved difficulty level to implement the autocomple feature can
range from moderate to difficult. The reason for it is that we have 3
different workflows, we have to work with file names and paths  as
well as get indexer support. Most of this complexity, however, comes
from the logic of the steps we have to do in order to
autocomplete. We get relatively easy access to information provided by
the components of \stexide{}. 

In the case of {\tt \tbs importmodule} command, the verification feature can
e seen as a particular case of autocompletion. Namely, it can safely
ignore the {\tt importTag} as nothing can go wrong there. For the
{\tt fileTag} it just has to check if a file with the specified name
exists, else add a corresponding error message. The {\tt idTag} should
get an Indexer factory instance, get the file index and query for
module IDs and finally check if the idTag corresponds to any of
them. Implementation is easier then but very similar to the one for
autocompletion.

The last implementation feature we will discuss in this section is the
indexing. Just as for the autocomplete feature we will have to
distinguish 3 cases depending on the tag we have to handle. If we get
the {\tt importTag}, we will set the property {\tt rdf:type
IDE:importModuleCommand} and will return true which means a new RDF
node will be created for it. For both the {\tt fileTag} and {\tt idTag} we
will analyze IPropertiesAcceptor stack and choose the first object
having the {\tt importTag}. Then will just add properties to it about
the file path and the module id. Considering the fact that we have
utility functions in \stexide{} for searching the stack for the first
AST of certain type, the method is quite short and easy to
understand. The complexity of the implementation can be set to easy.

\section{Conclusions}
\label{sec:discuss:conclusion}
I have presented the \stexide{} system, an integrated authoring
environment for \stex{} collections realized as a plugin to the
Eclipse IDE. While no extensible usability study was done, several
people (besides the author), already tried to use it and reported
positive feedback accompanied by some new feature requests. For now
the spectrum of new features to be implemented is still very broad
which is common for new IDEs. 

Looking back to the system implementation requirements (section
\ref{sec:sysimpl:req}), I think most 
of them were successfully implemented. The IDE architecture was
implemented in a modular way and the components which depend on \stex{} or
Eclipse were separated in different modules. In this way, a
good part of the architecture is \stex{} as well as Eclipse
independent. Hence implementation requirement I1 was achieved. Same
applies to requirement I2. Currently, the features responsible for
dealing with \stex{} specific tasks (e.g. handling theory definitions,
imports, symbol definitions) were implemented as extensions handled
through the handler registry. Any of them can be pluged-in or removed
at any time. A very important test for the whole plugin
mechanism was the fact that the aforementioned modules are very
interdependent e.g. to achieve autocompletion for module imports, one
has to be aware of what modules and symbols are defined. Due to the design
decisions I made, the interdependencies could be handled by the
generic tagging infrastructure and no implementation dependencies were
made. Both autocompletion and syntax highlighting features give user
some information about the context. While the syntax highlighting
gives a very rough estimation of the context, the autocompletion gives
very precise and localized information. That means requirement I3 was also
achieved. Through the validation mechanism I presented one can also
achieve some structure level redundancy detection. More complicated
redundancy checks will, in future, need a more sophisticated
infrastructure but I consider I4 to be 75\% achieved. Due to time
limitations, the requirement I5 was not implemented even though the
mechanism to generate the raw data is already working and is used by
the autocompletion and validation features. Hence only the User
Interface part is missing. The I6 implementation requirement was also
only partially achieved by the definition search which lay the basis
for other structural searches. More semantic searches, which is a
broad topic on its own, were not tackled.

Let us now examine whether the research questions (section
\ref{sec:aims:aims}) can be answered. 
\begin{compactenum}
\item [Question \textbf{A1}:] In the light of the experiences
  presented in \ref{sec:aims:motiv}, the semantic syntax highlighting,
  context-aware autocompletion and the future theory graph explorer
  make the process of flexiformalization more ``comfortable''. It is,
  still, a completely manual process and in the future work I will
  address the issue of making it semi-automatic. On the other hand, I
  expect that the features an IDE needs to make the flexiformalization process 
  easier will come as feature requests when the IDE becomes more
  popular. It makes a lot of sense to implement features as they get
  requested.
\item [Question \textbf{A2}:] As presented in section
  \ref{sec:arch:user_perspective}, I think the extension mechanism
  is a very important component to save implementation time costs for
  new features. I hope that in the future more people to write new
  extensions and increase the benefits one gets from flexiformalizing
  documents.
\item [Question \textbf{A3}:] As long as the context information is
  available in an explicit form, an IDE can provide it any
  user-friendly form. Hence providing context information depends more
  on the ability of the flexiformalization language (e.g. \stex{}) to
  markup this type of information. 
\item [Question \textbf{A4}:] In \stexide{} we have only very
  simple redundancy checks but not much support for creating reusable
  components. In the future we envision (section
  \ref{sec:discusion:feature_work}) the creation of several features
  to help the process of transforming an existing text into reusable
  components. 
\item [Question \textbf{A5}:] IDEs are the right tool to help the user
  work with highly interconnected structures. They can provide
  interactive ways to present, filter as well as reason about the
  connections. In \stexide{}, most of the features make use
  of the connections between content components -- hence, in some way,
  provide a natural way to interact with the components and links
  between them.  
\item [Question \textbf{A6}:] An IDE is as powerful as any
  other application running locally hence creating specialized
  mathematical indexes is just an engineering task. The benefit of
  IDEs is that they can make the search experience very interactive as
  well as combine it with flexiformalized structural
  information. Supporting math oriented semantic search is
  very important as it will be the most common way of discovering new
  content and reusing it.
\end{compactenum}

In conclusion, I would like to summarize the contributions of the
current research. From the conceptual point of view, this thesis
shed some light into the question of how to support flexiformalization
process so that it becomes as easy as possible. The implementation of
\stexide{}, laid the basis for future in-depth research in this
area. From system implementation point of view my main contribution is
the modular and extensible component architecture of an IDE geared
towards flexiformalization. This architecture was successfully
implemented and tested in \stexide{} but it is not bounded to either
\stex{} language or Eclipse framework. The same principles can be
reused for languages like \casl{}, LF\cite{Pfenning91} and in editors
like JEdit \cite{JEdit:web} etc. A
next contribution is the indexing infrastructure -- for its way of
orchestrating independent extensions towards creation of a consistent
picture of the document objects and their relations. A last system
implementation contribution is the Tagging mechanism which created a
common ground for communication between extensions while keeping
extension implementations completely independent. 

\section{Future Work}
\label{sec:discusion:feature_work}

There are several directions for future work. First is extending
\stexide{} with more features, for example:
\begin{compactenum}
  \item [\textbf{theory graph navigation}] will provide a graphical
    representation of the theory graph, give possibility to navigate
    to module definition by pressing the nodes of the graph.
  \item [\textbf{symbol presentation matching }] -- each \stex{}
    symbol definition introduces certain semantic meaning but in the same
    time specifies the way to render that symbol. For example, the
    \texttt{\tbs power} semantic macro has the expected mathematical
    meaning and specifies to be expanded into \{\#1\}\^\protect\{\#2\}
    where \#1 and \#2 are the first and second parameters to the
    macro. The symbol presentation matching feature will analyze the
    presentation of the semantic macro and when the user types
    x\^\protect2, it will suggest to transform it to {\texttt \tbs
      power\{x\}\{2\}}. One can also use this extension to search for
    all such flexiformalization possibilities in the whole corpora.
  \item [\textbf{module splitting}] -- a common case which showed up
    in previous attempts to flexiformalize documents
    \ref{sec:aims:motiv}, was that a certain theory was too big and
    needed to be splitted into several components. The problem was
    that the original theory imported other theories and by splitting
    it was not clear any longer which theories should be imported by
    which module. This can be easily handled by a module splitting
    feature which can also optimize it in a way that each resulting
    module imports the minimal number of other modules.
    \item [\textbf{outline filter}] -- a User Interface feature which
      was requested was to add filtering options to the document outline
      so that one can choose to see only very high level document
      structure (theory level). 
\end{compactenum} 

Other direction for the current research would be creating an IDE for
other languages like CASL, \omdoc{} or LF. Some people suggested
porting the architecture to their favourite editor like JEdit, Emacs
etc. 

\ifpdf
    \graphicspath{{7/figures/PNG/}{7/figures/PDF/}{7/figures/}}
\else
    \graphicspath{{7/figures/EPS/}{7/figures/}}
\fi



\chapter{Appendix}
  \section{Minimal \protect\stexide{} extension supporting syntax highlighting}
\label{minimal_syntaxhighlighting}
\begin{lstlisting}[language=java]
public class ImportModuleCommand implements IExtension {
  public static final String commandURI="kwarc.info.mkmide.latex.syntaxhighlighting.command";
  public static final String commandDesc="Command";

  public static final String externalRefURI="kwarc.info.mkmide.latex.syntaxhighlighting.externalRef";
  public static final String externalRefDesc="External references";

  public static final String importModuleCommandTag="kwarc.info.mkmide.latex.importmodule.commandtag";
  public static final String importModuleFileTag="kwarc.info.mkmide.latex.importmodule.filetag";
  public static final String importModuleSymbolTag="kwarc.info.mkmide.latex.importmodule.symboltag";

  static final String [] tags = {importModuleCommandTag, importModuleFileTag, importModuleSymbolTag}; 
	
  static final String HandledCommandNames[] = {"importmodule"};

  public String[] getHighlightingURIs() {
    return new String [] { commandURI, externalRefURI };
  }

  public String[] getHandledCommandNames() {
    return HandledCommandNames;
  }

  public void addNodeTags(Command cmd) {
    if (cmd.getOptions().size()==0)
      return;
    Options option = cmd.getOptions().get(0);
    cmd.getCommandTags().add(importModuleCommandTag);
    if (cmd.getOptions().size()==1)
      return;		
    option = cmd.getOptions().get(1);
    recursivelyApplyTag(option, importModuleSymbolTag);
  }

  public String[] getHandledTags() {
    return tags;
  }

  public String getSyntaxColorURI(String tag) {
    if (importModuleFileTag.equals(tag) || importModuleSymbolTag.equals(tag)) {
      return externalRefURI;
    } else
    if (importModuleCommandTag.equals(tag)) {
      return commandURI;
    }
    return null;
  }

  public String getDescription(String highlightingURI) {
    if (commandURI.equals(highlightingURI)) {
      return commandDesc;
    } else
    if (externalRefURI.equals(highlightingURI)) {
      return externalRefDesc;
    }
    return null;
  }
}
\end{lstlisting}








\bibliographystyle{alpha} 
\renewcommand{\bibname}{References} 

\bibliography{9_backmatter/kwarc} 

\printindex








\end{document}